\documentclass[twocolumn, secnumarabic, amssymb, amsmath,nobibnotes, aps, prl,nofootinbib]{revtex4}
\usepackage{graphics}

\begin{document}
\title{hyperbolic nature of uniformly rotating systems and their relation to gravity}
\author{B. H. Lavenda}
\email{bernard.lavenda@unicam.it}
\affiliation{Universit$\grave{a}$ degli Studi, Camerino 62032 (MC) Italy}
\date{\today}
\newcommand{\sumn}{\sum_{i=1}^{n}\,}
\newcommand{\sumk}{\sum_{i=1}^{k}\,}
\newcommand{\half}{\mbox{\small{$\frac{1}{2}$}}}
\newcommand{\fourth}{\mbox{\small{$\frac{1}{4}$}}}
\newcommand{\twothirds}{\mbox{\small{$\frac{2}{3}$}}}
\newcommand{\third}{\mbox{\small{$\frac{1}{3}$}}}
\newcommand{\fifth}{\mbox{\small{$\frac{1}{5}$}}}
\newcommand{\sixth}{\mbox{\small{$\frac{1}{6}$}}}
\newcommand{\forty}{\mbox{\small{$\frac{3}{40}$}}}
\newcommand{\ten}{\mbox{\small{$\frac{3}{10}$}}}
\newcommand{\eighth}{\mbox{\small{$\frac{1}{8}$}}}
\newcommand{\fourthirds}{\mbox{\small{$\frac{4}{3}$}}}
\newcommand{\threefourths}{\mbox{\small{$\frac{3}{4}$}}}
\newcommand{\nn}{\mbox{\small{$\frac{1}{n}$}}}
\newcommand{\lc}{\lambda_{\mbox{\tiny C}}}
\newcommand{\lt}{\lambda_{\mbox{\tiny T}}}
\newcommand{\ob}{\overline{\beta}}
\newcommand{\rr}{\overline{r}}
\begin{abstract}
Special relativity corresponds to hyperbolic geometry at constant velocity while the so-called general relativity corresponds to hyperbolic geometry of uniformly accelerated systems. Generalized expressions for angular momentum, centrifugal and Coriolis forces are found in hyperbolic space, which reduce to the usual expressions of Euclidean space when the absolute constant tends to infinity. Gravity enters only in the specification of  the absolute constant. A uniformly rotating disc corresponds exactly to hyperbolic geometry with a constant negative Gaussian curvature. The angle defect is related to Lorentz contraction of objects normal to the radial direction.  Lobachevsky's angle of parallelism accounts for the apparent relativistic distortion of moving objects and would provide a testing ground  to measure a positive defect by replacing large distances by high speeds that are comparable with that of light.
\end{abstract}
 
\maketitle

\section{relativity in a hyperbolic world}

General relativity was conceived to treat the equivalence of observations made by accelerated observers~\cite{Bondi}. Consider a rotating disk of constant angular velocity $\omega$ and radius $\rr$. At the center of the disk we have an inertial system which is described by Euclidean geometry. It is well-known~\cite{Moller} that a clock located anywhere else on the disk will have a velocity $\rr\omega$ relative to the inertial system and its clock will be retarded by the amount:
\[
\tau=t\sqrt{1-\rr^2\omega^2/c^2}. \]

Now, it is argued that any rod in motion should undergo a Lorentz contraction. This means that any two points on the disk that are at a distance $\rr$ from the center, say, $\rr,\theta$ and $\rr,\theta+d\theta$ should have a rod which is shortened with respect to the length of the rod in the inertial frame $d\rr_0$ by an amount:
\[
\rr\,d\theta=d\rr_0\sqrt{1-\rr^2\omega^2/c^2}. \]
It therefore follows the length of the circumference of the circle will be greater than if the disk were stationary:
\[
\int_0^{2\pi}\frac{\rr\,d\theta}{\sqrt{1-\rr^2\omega^2/c^2}}>\rr\int_0^{2\pi}d\theta=2\pi\rr.
\]

It is well-known that by replacing the radius by $i$ times the \lq radius\rq, formulas for circumference of hyperbolic circles, areas and volumes are greater than their Euclidean counterparts. Here, we  compare the non-flat metrics of general relativity that describe curved space and time with the simpler hyperbolic differential of arc length which is a product of the flat metric for Euclidean space and a differential coefficient that behaves in much the same way as a varying index of refraction in an inhomogeneous medium. Such an approach has been used previously to derive the three celebrated tests of general relativity from a flat spatial metric~\cite{Lavenda}. 

Moreover, the logarithmic definition of hyperbolic distance, through the cross ratio, is a generalized expression for the Doppler shift. The geodesics derived from the hyperbolic metric will show that the ratio of the Euclidean measure of the speed to the speed of light is given by the cosine of the angle of parallelism. This will explain the distortion of objects not as a Lorentz contraction,  but,  as an angular defect like the annual oscillation of a star's apparent position due to the earth's motion about the sun. Since the angle of parallelism is an upper bound on the angle defect of a hyperbolic triangle, it may offer support to the idea that the physical space of relativity is actually Lobachevskian rather than Euclidean or spherical. The space constant rather than being a measure of the curvature of space, which may be arbitrarily large, contains the finite velocity of light which is no limitation on the hyperbolic  measure of velocity.

\section{Analogy with Fermat's principle of least time}
According to Fermat's principle the ray path connecting two arbitrary points, $[t_1,t_2]$, in the plane $(x_1,x_2)$, make the optical path length
\begin{equation}
s=\int_{t_{1}}^{t_{2}}\eta(x,\dot{x})\sqrt{\dot{x}_1^2+\dot{x_2}^2}dt, \label{eq:Fermat}
\end{equation}
stationary,
where $\eta$ is the index of refraction which may depend on position $x$, as well as velocity $\dot{x}$, and the dot denotes differentiation with respect to coordinate time $t$.  Expanding the logarithmic expression for the distance in terms of Beltrami coordinates, the hyperbolic differential of the arc length determines the index of refraction as:
\begin{equation}
\eta(x,\dot{x})=\kappa\frac{\sqrt{1-\left(x_1\dot{x}_2-x_2\dot{x}_1\right)^2/\left(\dot{x}_1^2+\dot{x}_2^2\right)}}{1-x_1^2-x_2^2}, \label{eq:eta}
\end{equation}
where $\kappa$ is known as the radius of curvature or space constant, being the distance between concentric limiting arcs whose ratio is $e$. Although it is an example of an absolutely determined length, its numerical value will depend upon the arbitrary choice of the unit of measurement.

Consequently, the hyperbolic line element becomes:
\begin{eqnarray}
ds & = &\eta(x,\dot{x}) d\bar{s}\nonumber\\
& = & \kappa\frac{\sqrt{\dot{x}_1^2+\dot{x}_2^2-\left(x_1\dot{x}_2-x_2\dot{x_1}\right)^2}}{1-x_1^2-x_2^2}dt, \label{eq:hyper}
\end{eqnarray}
where 
\begin{equation}
d\bar{s}=\sqrt{dx_1^2+dx_2^2} \label{eq:euclid}
\end{equation}
is the Euclidean differential of the arc length. The true nature of the index of refaction is that it is the differential of the hyperbolic arc length with respect to its Euclidean measure.

By transforming to polar coordinates, $x_1=(\rr/\kappa)\cos\theta$ and $x_2=(\rr/\kappa)\sin\theta$, the  physical significance of the terms in (\ref{eq:hyper}) become apparent. For then we have
\[
ds=\frac{\sqrt{\dot{\rr}^2+\rr^2\dot{\theta}^2-(\rr^2\dot{\theta})^2/\kappa^2}}{1-\rr^2/\kappa^2}\,dt. \]
The first two terms in the numerator is twice the kinetic energy, 
\begin{equation}
2\bar{T}=\dot{\rr}^2+\rr^2\dot{\theta}^2, \label{eq:T}
\end{equation}
while the last term is the square of the angular momentum, 
\begin{equation}
L=\rr^2\dot{\theta}. \label{eq:L}
\end{equation}

We can write (\ref{eq:hyper}) alternatively as:
\begin{equation}
ds=\frac{\sqrt{d\rr^2+\rr^2d\theta^2\left(1-\rr^2/\kappa^2\right)}}{1-\rr^2/\kappa^2}. \label{eq:lob}
\end{equation}
For $\theta=\mbox{const.}$, (\ref{eq:lob}) reduces to
\begin{equation}
ds=\frac{d\rr}{1-\rr^2/\kappa^2}, \label{eq:lob1}
\end{equation}
which can easily be integrated to give
\[
s=\frac{\kappa}{2}\ln\left(\frac{1+\rr/\kappa}{1-\rr/\kappa}\right)=\kappa\tanh^{-1}(\rr/\kappa)=r. \;\;\;\;\;\;\theta=\mbox{const.}
\]
This shows that at constant $\theta$, the Euclidean measure of length, $\rr$, is related to the corresponding segment of the Lobachevsky straight line by
\begin{equation}
\rr=\kappa\tanh(r/\kappa). \label{eq:tanh}
\end{equation}

Next consider $\rr=\mbox{const.}$ Then (\ref{eq:lob}) reduces to
\begin{equation}
ds=\frac{\rr d\theta}{\sqrt{1-\rr^2/\kappa^2}}. \label{eq:lob2}
\end{equation}
Integrating from $0$ to $2\pi$ gives:
\begin{equation}
s=\frac{2\pi\rr}{\sqrt{1-\rr^2/\kappa^2}}>2\pi\rr, \;\;\;\;\;\;\;\rr=\mbox{const.} \label{eq:lob-circle}
\end{equation}
which shows that the length of the circumference of a hyperbolic circle of radius $r$ is greater than that of a Euclidean circle of radius $\rr$. For upon introducing (\ref{eq:tanh}) into (\ref{eq:lob-circle}) we get: 
\[
s=2\pi\kappa\sinh(r/\kappa),\]
which is the well-known expression for the circumference of a hyperbolic circle of radius $r$~\cite{Busemann}.

The general expression for the square of the hyperbolic arc length is
\begin{equation}
ds^2=\kappa^2\left[\frac{d\rr^2}{(1-\rr^2/\kappa^2)^2}+\frac{\rr^2d\theta^2}{1-\rr^2/\kappa^2}\right]. \label{eq:hyper-2}
\end{equation}
Expression \eqref{eq:hyper-2} is comparable with the expression given by general relativity~\cite[p. 224]{Moller}:
\begin{equation}
ds^2=dr^2+\frac{r^2d\theta^2}{1-r^2\omega^2/c^2}. \label{eq:gr-sigma}
\end{equation}
Although it was recognized by general relativitists that the geometry is no longer Euclidean, no connection was made with hyperbolic geometry. In fact, the first term in \eqref{eq:gr-sigma} is incorrect since at constant $\theta$ it does not integrate to give the hyperbolic measure of the radius, but, rather, it gives the Euclidean measure. The spatial scaling constant is $\kappa=c/\omega$, while the velocity scaling is $\kappa=c$.

\section{geodesic paths}

 Choosing $\rr$ as the independent variable equation, the differential of the hyperbolic arc length (\ref{eq:hyper}) can be written as:
\begin{eqnarray}
ds & = & s^{\prime}d\rr\nonumber\\
& = &\kappa\frac{\sqrt{	1+(\rr\theta^{\prime})^2\left(1-\rr^2/\kappa^2\right)}}{1-\rr^2/\kappa^2}d\rr, \label{eq:hyper-bis}
\end{eqnarray}
where the prime stands for differentiation with respect to the independent variable, $\rr$.
Since $\theta$ is a cyclic coordinate, (\ref{eq:hyper-bis}) immediately gives a first integral,
\begin{equation}
\kappa\frac{\partial s^{\prime}}{\partial\theta^{\prime}}=\frac{\rr^2\theta^{\prime}}
{\sqrt{1+\left(\rr\theta^{\prime}/\kappa\right)^2(1-\rr^2/\kappa^2})}=L/c=\mbox{const.} \label{eq:L-bis}
\end{equation}
Solving (\ref{eq:L-bis}) for $\theta^{\prime}$ gives
\begin{equation}
\theta^{\prime}=\frac{\dot{\theta}}{\dot{\rr}}=\pm\frac{L/c}{\rr^2\sqrt{1-(L^2/c^2\rr^2)\left(1-\rr^2/\kappa^2\right)}}, \label{eq:theta}
\end{equation}
where the prime stands for differentiation with respect to $\rr$. Integrating \eqref{eq:theta} gives
\begin{equation}\cos\theta=\frac{a}{\rr},\label{eq:polar}
\end{equation}
where $a=L/c\sqrt{1-\rr^{2}/\kappa^{2}}$. The particle will travel in a straight line since \eqref{eq:polar} is its polar normal form.

What we have accomplished is to find the geodesic curve, which can be written more generally as:
\[
\theta^{\prime}=\pm\frac{(L/c)\sqrt{E}}{\sqrt{G}\sqrt{G-(L/c)^2}}, \]
where the metric coefficients in
\[ds^2=E\;d\rr^2+G\;d\theta^2,\]
are $E=1/(1-\rr^2/\kappa^2)^{2}$ and $G=\rr^2/(1-\rr^2/\kappa^2)$. The Gaussian curvature can be expressed in terms of the metric coefficients; in the present case it is given by:
\begin{equation}
K=-\frac{1}{2\sqrt{EG}}\frac{d}{d\rr}\left(\frac{G_{\rr}}{\sqrt{EG}}\right). \label{eq:K}
\end{equation}The geodesic equations follow from differentiating
\begin{eqnarray}
\dot{\rr}&=&\frac{\sqrt{G-(L/c)^2}}{\sqrt{GE}}, \nonumber\\
&=& c\left(1-\rr^{2}/\kappa^{2}\right)\sqrt{1-\left(L^{2}/c^{2}\rr^{2}\right)\left(1-\rr^{2}/\kappa^{2}\right)}\label{eq:r-geo}
\end{eqnarray}
and
\begin{equation}
G\dot{\theta}=\frac{\rr^2\dot{\theta}}{1-\rr^2/\kappa^2}=L=\mbox{const.}, \label{eq:theta-geo}
\end{equation}
with respect to $t$. We then obtain the well-known (geodesic)  equations of motion:
\begin{equation}
\ddot{\rr}+\frac{1}{2E}\left(E_{\rr}\dot{\rr}^2-G_{\rr}\dot{\theta}^2\right)=0
\label{eq:r-eom}
\end{equation}
and
\begin{equation}
\ddot{\theta}+\frac{G_{\rr}}{G}\dot{\rr}\dot{\theta}=0, \label{eq:theta-eom}
\end{equation}
where the subscript denotes differentiation with respect to the variable.

As \eqref{eq:theta-geo} clearly shows, it is necessary to modify the definition of the angular momentum in curved space. M{\o}ller \cite{Moller} assumes that $(1-\rr^2/\kappa^2)^{-1}$ is a small correction to the angular momentum, and calculates the correction term for Mercury. Rather, we argue that \eqref{eq:theta-geo} is exact for hyperbolic space. 

In Euclidean space,
\begin{equation}
\ddot{\rr}=\rr\dot{\theta}^2,\;\;\;\;\;\;\;\;\;\; \rr\ddot{\theta}=-2\dot{\rr}\dot{\theta}=\frac{L^2}{\rr^3} \label{eq:a-Euclid}
\end{equation}
are the centrifugal and Coriolis accelerations, respectively. Notice that the velocity of light has disappeared. In contrast, in hyperbolic space they become
\begin{equation}
\ddot{\rr}=\rr\dot{\theta}^2-\frac{2\rr}{\kappa^2}\frac{\dot{\rr}^2}{1-\rr^2/\kappa^2}, \label{eq:a-hyp-r}
\end{equation}
and
\begin{equation}
\rr\ddot{\theta}=-2\frac{\dot{\rr}\dot{\theta}}{\-\rr^2/\kappa^2}, \label{eq:a-hyper-theta}
\end{equation}
respectively.  These equations insure that the acceleration of the curve is normal to the surface, and perpendicular to the velocity. For a circular orbit, $\dot{\rr}=0$, and the angular velocity attains its maximum value
\[\rr\dot{\theta}=c\sqrt{1-\rr^2/\kappa^2}.\]
General relativity predicts \eqref{eq:theta-geo} and \eqref{eq:a-hyper-theta}, but claim that the conservation of angular momentum is violated, but only slightly \cite[eqn (76a) p. 241]{Moller}.
Introducing \eqref{eq:theta} into \eqref{eq:lob} gives
\begin{equation}
ds^{\star}=\frac{d\rr}{(1-\rr^2/\kappa^2)\sqrt{1-(L^2/c^2\rr^2)(1-\rr^2/\kappa^2)}}. \label{eq:lob-star}
\end{equation}
 Were it not for the factor $(1-\rr^2/\kappa^2)$ in the denominator, we could set $ds^{\star}=cdt$ so as to fulfill Fermat's principle. This would relate Fermat's principle to Euclidean, or coordinate, time, and not to hyperbolic, or proper, time.

 The calculation of the distance $s_{12}$ between points $r_{1}$ and $r_{2}$,
 \begin{equation} 
 s_{12}=\int_{\rr_{1}}^{\rr_{2}}\frac{\sqrt{d\rr^{2}+\rr^{2}d\theta^{2}\left(1-\rr^{2}/\kappa^{2}\right)}}
 {1-\rr^{2}/\kappa^{2}} \label{eq:s12}
 \end{equation}
 conforms to the Poincar\'e model.  To this end, we introduce the effective centrifugal potential
 \[V_{E}=\frac{L^{2}}{2\rr^{2}}\left(1-\frac{\rr^{2}}{\kappa^{2}}\right),\]
 where $\kappa=c/\sqrt{G\varrho}Ä$, for a mass of constant density $\varrho$, and $G$ is the Newtonian gravitational constant. This choice of the absolute constant sets the Gaussian curvature \eqref{eq:K} directly proportional to the constant mass density, viz., $K=-1/\kappa^{2}=-G\varrho/c^{2}$.
 Squaring \eqref{eq:r-geo} gives the energy conservation law,
 \[\half\frac{\dot{\rr}^{2}}{\left(1-\rr^{2}/\kappa^{2}\right)^{2}}+V_{E}=c^{2}/2.\]
 Integrating the definite integral \eqref{eq:s12} results in 
 \begin{eqnarray*}
 \lefteqn{s_{12}= \frac{\kappa}{2}\times}\\
 & &\ln\left(\frac{\kappa+\rr_{2}\sqrt{1-2V_{E}(\rr_{2})/c^{2}}}{\kappa-\rr_{2}
	 \sqrt{1-2V_{E}(\rr_{2})/c^{2}}}\cdot\frac{\kappa-\rr_{1}\sqrt{1-2V_{E}(\rr_{1})/c^{2}}}{\kappa+\rr_{1}\sqrt{1-2V_{E}(\rr_{1})/c^{2}}}\right).
	 \end{eqnarray*}
For large radial velocities, or equivalently, at low angular momentum, the distance between points $r_{1} $ and $r_{2}$ becomes the Poincar\'e length
\begin{equation}
s_{12}=\frac{\kappa}{2}\ln R, \label{eq:P}
\end{equation}
where $R$ is the cross-ratio~\cite[p. 177]{Busemann}:
\begin{equation}
R=\frac{\kappa+\rr_{2}}{\kappa-\rr_{2}}\cdot\frac{\kappa-\rr_{1}}{\kappa+\rr_{1}}, \label{eq:cr}
\end{equation}
between the ordered points $(\kappa,\rr_{2},\rr_{1},-\kappa)$. The Poincar\'e length \eqref{eq:P} length vanishes when $r_{1}=r_{2}$ and tends to infinity when either $\rr_{2}\uparrow\kappa$ or $\rr_{1}\downarrow-\kappa$, where $\kappa$ is the radius of the Poincar\'e semi-circle. This is to say that the distance to the Cartesian radius $\kappa$ is infinite! This also illustrates the important fact that gravitational considerations enter only through the specification of the ideal points, in this case the absolute constant, $\kappa$.

\section{hyperbolic geometry}
There are actually two scaling parameters in (\ref{eq:lob-star}). The time dilatation factor is:
\begin{equation}
\sqrt{1-\dot{\rr}^2/c^2}=(\rr\dot{\theta}/c)\sqrt{1-\rr^2\omega^2/c^2}, \label{eq:di}
\end{equation}
where $\rr\omega$ is the velocity of the clock relative to the inertial system located at the origin of the disk. It is evident from \eqref{eq:di} that the dilatation of rectilinear motion is actually a decrease in the angular velocity of an object, $\rr\dot{\theta}$, on the disk due to the angular velocity of rotation of the disk, $\rr\omega$.

Consider the triangle inscribed in a circle of radius $1$ in the figure. If the Euclidean measure of the  sides are $\bar{\beta}=\dot{\rr}/c$ and $\bar{\alpha}=\rr\dot{\theta}\sqrt{1-\rr^2\omega^2/c^2}/c$,  then the hypotenuse $\bar{\gamma}=1$. The angle $A$ at the center of the disk will obey Euclidean geometry so that its Euclidean measure $\bar{A}$ will coincide with its hyperbolic measure $A$:
\begin{eqnarray*}
\cos A & = & \cos\bar{A}=\bar{\beta}\\
& = & \dot{\rr}/c=\tanh(\beta/\kappa)=\tanh(\dot{r}/c), 
\end{eqnarray*}
since we are at the limit $\gamma\rightarrow\infty$ and $\bar{\gamma}=\tanh(\gamma/\kappa)=1$. 

However, because the angle $B$ is non-central, it Euclidean, $\bar{B}$, and hyperbolic, $B$, measures will not coincide. Denoting $d(x,y)$ as the Euclidean distance between $x$ and $y$, the logarithm of the cross ratio is the length in hyperbolic space:
\begin{eqnarray*}
\alpha & = & \frac{\kappa}{2}\ln\left(\frac{d(c,u)}{d(b,u)}\cdot\frac{d(b,v)}{d(c,v)}\right)\\
& = & \frac{\kappa}{2}\ln\left[\frac{\mbox{sech}(\beta/\kappa)}{\mbox{sech}(\beta/\kappa)-\bar{\alpha}}\cdot
\frac{\mbox{sech}(\beta/\kappa)+\bar{\alpha}}{\mbox{sech}(\beta/\kappa)}\right]. 
\end{eqnarray*}
However, 
\[\bar{\alpha}=\mbox{sech}\beta/\kappa=\sqrt{1-\bar{\beta}^2/\kappa^2},\] so that the denominator in the first term in the argument of the logarithm will vanish implying $\alpha=\infty$. Hence,
\begin{equation}
\bar{\alpha}=\rr\dot{\theta}\sqrt{1-\rr^2\omega^2/c^2}/c=\mbox{sech}(\beta/\kappa)=\mbox{sech}(r\dot{\theta}/c). \label{eq:alpha}
\end{equation}

Now, the cosine of the hyperbolic measure of the angle $B$ will be $\cos B=\tanh(\alpha/\kappa)/\tanh(\gamma/\kappa)$, or the ratio of the adjacent to the hypotenuse. Since both $\alpha$ and $\gamma$ are both infinite, the ratio of their hyperbolic tangents will be $1$, and hence 
\begin{eqnarray}
\cos\bar{B} & = & \bar{\alpha}/\bar{\gamma}=\frac{\tanh(\alpha/\kappa)}{\tanh(\gamma/\kappa)}\mbox{sech}(\beta/\kappa)\nonumber\\
& = & \cos B\;\mbox{sech}(\beta/\kappa)=\mbox{sech}(\beta/\kappa). \label{eq:B}
\end{eqnarray}
Since $\mbox{sech}(\beta/\kappa)$ is the greatest value of $\alpha$, it follows that $\bar{B}>B$ on the interval $(0,\pi)$ where the cosine is a decreasing function. And since $\bar{B}=\pi/2-A$, the sum of the angles of a hyperbolic triangle will be less than $\pi$, which is the well-known angle defect. 

The Euclidean Pythagorean theorem, $\bar{\gamma}^2=\bar{\alpha}^2+\bar{\beta}^2$, instead of leading to the well-known hyperbolic Pythagorean theorem:
\begin{equation}
\cosh(\gamma/\kappa)=\cosh(\alpha/\kappa)\cosh(\beta/\kappa),\label{eq:Pyth}
\end{equation}
 gives, rather, the trigonometric identity:
\[1=\tanh^2(\dot{r}/c)+\mbox{sech}^2(\dot{r}/c).\]

Since the angle $A$ is at the center of the disc, its hyperbolic and Euclidean measures will coincide:
\[\cos A=\cos\bar{A}=\frac{\dot{\rr}}{\sqrt{2\bar{T}}}=\frac{\tanh(\dot{r}/c)}{\tanh(\sqrt{2T}/c)},\]
and
\begin{equation}
\rr\dot{\theta}=\tanh(r\dot{\theta}/c)\mbox{sech}(\dot{r}/c),\label{eq:shrink}
\end{equation}
so that
\begin{eqnarray*}
\cos\bar{B} & = & \frac{\rr\dot{\theta}}{\sqrt{2\bar{T}}}=\frac{\tanh(r\dot{\theta}/c)}{\tanh(\sqrt{2T}/c)}\mbox{sech}(\dot{r}/c)\\
& = &\cos B\sqrt{1-\dot{\rr}^2/c^2}.\end{eqnarray*}
From this it follows that $\bar{B}>B$:  \emph{This is the relativistic origin of the angle defect in hyperbolic triangles\/}.  

Finally, the hyperbolic Pythagorean theorem \eqref{eq:Pyth} is given explicitly as:
\[\cosh(\sqrt{2T}/c)=\cosh(r\dot{\theta}/c)\cdot\cosh(\dot{r}/c).\]

Due to the rotation of the disk, the side $\alpha$ will be shortened and the angle $B$ smaller than if the disk were at rest. Squaring both sides of (\ref{eq:shrink}) gives the equation of an ellipse
\begin{equation}
\left(\frac{\rr\dot{\theta}}{c}\right)^2\mbox{coth}^2(r\dot{\theta}/c)+\dot{\rr}^2/c^2=1. \label{eq:ellipse}
\end{equation}
In the limit as $r\rightarrow\infty$, the ellipse (\ref{eq:ellipse}) degenerates into a circle. Again, this is not the equation one would get from general relativity~\cite[Eqn (76b) on p. 241]{Moller}
\[\left(\frac{\dot{r}}{c}\right)^2+\frac{1}{1-r^2\omega^2/c^2}
\left(\frac{r\dot{\theta}}{c}\right)^2=1,\]
which violates the conservation of energy.

\section{hyperbolic nature of the electromagnetic field and the Poincar\'{e} stress}

Consider a charge moving in the $x$-direction at a constant speed $c\beta$. The law of transformation of the electromagnetic fields, $E$ and $H$, are:
\begin{subequations}
\begin{align}
E^{\prime}_{x}=E_{x}\;\;\;\;\;\;\;\;\;\;\;\;\;\;\;\;\;\;\;\; H^{\prime}_{x}=H_{x}\label{eq:M1}\\
E^{\prime}_{y}=\Gamma\left(E_{y}-\beta H_{z}\right)\;\;\;\;\;\;\;\;\;\;\;\; H^{\prime}_{y}=\Gamma\left(H_{y}+\beta E_{z}\right)\label{eq:M2}\\
E^{\prime}_{z}=\Gamma\left(E_{z}+\beta H_{y}\right) \;\;\;\;\;\;\;\;\;\;\;\; H^{\prime}_{z}=\Gamma\left(H_{z}-\beta E_{y}\right),\label{eq:M3}
\end{align}
\end{subequations}
where $\Gamma=1/\sqrt{1-\beta^{2}}$. Furthermore, consider the prime inertial system to be at rest, and, for continuity with the previous section consider the $xy$ plane. 

According to hyperbolic geometry the sides may be expressed in terms of the angles. Consequently, the first two transformation laws, \eqref{eq:M1} and \eqref{eq:M2} can be stated as:
\begin{subequations}
\begin{align}
\cos A  = \cos\bar{A}=\bar{\beta}/\bar{\gamma}=\tanh\beta/\tanh\gamma\;\;\;\;\;\;\;\;\;\;\;\;\;\;\; \label{eq:M1-bis}\\
\mbox{sech}\beta\cos B=\cos\bar{B}
=\bar{\alpha}/\bar{\gamma}\;\;\;\;\;\;\;\;\;\;\;\;\;\;\;\;\;\;\nonumber\\
\;\;\;\;\;\;\;\;\;\;\;\;=\mbox{sech}{\beta}\tanh{\alpha}/\tanh{\gamma}=\sin A=\sinh\alpha/\sinh\gamma\;\;\;\;\;\;\;\;\;\;\;\; \label{eq:M2-bis}
\end{align}
\end{subequations}
where the latter follows from the fact that $H^{\prime}_{z}=0$, and it is an expression of the \emph{hyperbolic\/} Pythagorean theorem \eqref{eq:Pyth}.

It is well-known that the components of the force are obtained by multiplying their projections in the $x$, $y$, and $z$ planes by the factors $1$, $\Gamma^{-1}$, and $\Gamma^{-1}$, respectively. In the $xy$ plane the force components will be given by:
\begin{subequations}
\begin{align}
F_{x}=\frac{e}{4\pi r^{2}}E_{x}=\frac{e}{4\pi r^{2}}\cos\bar{A} \label{eq:F1}\\
F_{y}=\frac{e}{4\pi r^{2}}\left(E_{y}-\beta H_{z}\right)\nonumber\\
=\frac{e}{4\pi r^{2}}E^{\prime}_{y}/\Gamma=\frac{e}{4\pi r^{2}}\cos B\mbox{sech}\beta, \label{eq:F2}
\end{align}
\end{subequations}
where $r$ is the radius of the sphere of charge $e$.  The magnitude of the force being
\begin{equation}
 F=\sqrt{F_{x}^{2}+F_{y}^{2}}=\frac{e}{4\pi r^{2}}\sqrt{\cos^{2}\bar{A}+\cos^{2}\bar{B}}. \label{eq:F}
 \end{equation}
 Without realizing that
 \begin{eqnarray*}
 \lefteqn{\cos^{2}\bar{A}+\cos^{2}\bar{B}=}\\
 & &\frac{1-\mbox{sech}^{2}\beta+(1-\mbox{sech}^{2}\alpha)\mbox{sech}^{2}\beta}{\tanh^{2}\gamma}=1\end{eqnarray*}
 precisely on account of the hyperbolic Pythagorean theorem, \eqref{eq:Pyth},  Page and Adams÷\cite{Page} invent a charge conservation per unit area on the surface of the electron, $\rho d\sigma=\rho^{\prime}d\sigma^{\prime}$, where $\rho^{\prime}=e/4\pi r^{2}$, and the surface elements are supposedly related by:
 \[d\sigma=d\sigma^{\prime}\sqrt{\cos^{2}\bar{A}+\cos^{2}\bar{B}}.\]
 Then solving for the unknown $\rho$, they eliminate the square root in \eqref{eq:F}. Since the electromagnetic field vanishes inside the electron, the stress acting on the surface is:
 \[
 \mathcal{S}=\half F^{2}=\frac{e^{2}}{32\pi^{2} r^{4}},\]
 which is the well-known Poincar\'{e} stress that was needed to reduce the $4/3$ factor in the expression for the energy to unity, and led to the conclusion that the mass of an electron was not totally electromagnetic in origin.

\section{terrell-weinstein effect and the angle of parallelism}

If we want to determine the size of a rod traveling at a velocity $\dot{\rr}$, the photons we observe emanating from the ends of the rod will arrive at different times. Terrell~\cite{Terrell} showed that one can interpret what is usually viewed as a FitzGerald-Lorentz contraction as a distortion due to the rotation of the rod. Weinstein~\cite{Weinstein} claimed, about the same time, that the length of a rod can appear infinite which he claimed cannot be due to a mere rotation. Here, we will show it to be due a phenomenon analogous to stellar parallax, and involves  the angle of parallelism in hyperbolic geometry.

In the limiting case we have the Euclidean distance $d(0,c)=\cos A=\dot{\rr}/c$ (see the figure, since the hypotenuse is $1$. Thus, the hyperbolic measure of the velocity, $\dot{r}$, whose Eulcidean value satisfies $\dot{\rr}/c<1$, is:
\begin{eqnarray}
\dot{r}/c & = & \half\ln\left(\frac{1+\dot{\rr}/c}{1-\dot{\rr}/c}\right)\nonumber\\
& = & \half\ln\left(\frac{1+\cos A}{1-\cos A}\right)=\half\left(\frac{1+\cos A}{\sin A}\right)^2\nonumber\\
 & = & \ln\mbox{cot}(A/2),\label{eq:dis}
\end{eqnarray}
where we have used a half-angle trigonometric formula. 

The angle $A$ is called the parallel-angle, $\Pi(\dot{r})$, and it is a sole function of $\dot{r}$. Exponentiating both sides of \eqref{eq:dis} results in the celebrated Lobachevsky formula:
\begin{equation}
\mbox{cot}(\Pi\left(\dot{r})/2\right)=e^{\dot{r}/c}. \label{eq:Lobachevsky}
\end{equation}

Consider a rod moving with velocity $\dot{\rr}$ along the $r$ axis. Light from the trailing and leading edges must travel over different distances, and, hence arrive a different times. Suppose the distance covered by photons emanating from the trailing edge is $d_1$, while that from the leading edge $d_2$. Their respective times are $d_1/c$ and $d_2/c$. The observer that sees light at time $t$ will have emanated from the trailing and leading edges at $t-d_1/c$ and $t-d_2/c$, respectively, because of the finite propagation of light. 

The Lorentz transformations for the space coordinates will then be:
\begin{subequations}
\begin{align}
r_1^{\prime}=\Gamma\left[r_1-\dot{\rr}(t-d_1/c)\right]\nonumber\\
r_2^{\prime}=\Gamma\left[r_2-\dot{\rr}(t-d_2/c)\right],\nonumber\\
\end{align}
\end{subequations}
where $\Gamma=1/\sqrt{1-\dot{\rr}^2/c^2}$.
Their difference provides a relation between the lengths of the rod in the system traveling at the velocity $\dot{\rr}$, $\ell^{\prime}=r_2^{\prime}-r_1^{\prime}$, and that measured in the coordinate system at rest, $\ell=r_2-r_1$, which is:
\[
\ell^{\prime}=\Gamma\left[\ell+(\dot{\rr}/c)(d_2-d_1)\right].\]
Now, the difference in distances travelled by the photons from the leading and trailing edges is just the length of the rod in the system at rest, and so we have:
\begin{equation}
\ell^{\prime}=\ell\left(\frac{1+\dot{\rr}/c}{1-\dot{\rr}/c}\right)^{1/2},\label{eq:Doppler}
\end{equation}
if the rod is approaching the stationary observer. For a  rod receding from the observer, the signs in the numerator and denominator must be switched because $\dot{\rr}\rightarrow-\dot{\rr}$. We could have arrived at \eqref{eq:Doppler} directly by observing that in addition to the usual Doppler effect there is a time dilatation between observers located on the moving and stationary frames.

Setting $\dot{\rr}/c$ equal to the cosine of the angle of parallelism in the Doppler expression \eqref{eq:Doppler} gives
\begin{equation}
\ell^{\prime}/\ell=\cot\left(\Pi(\dot{r})/2\right)=e^{\dot{r}/c}, \label{eq:approach}
\end{equation}
if the rod is approaching, while
\begin{equation}
\ell^{\prime}/\ell=\tan\left(\Pi(\dot{r})/2\right)=e^{-\dot{r}/c}, \label{eq:recede}
\end{equation}
if it is receding. The angle of parallelism must be greater than $A$, and the larger the relative velocity $\dot{\rr}/c$ the smaller will be the angle $A$.  Thus, we would expect to see a large expansion of the object as it approaches us, and a corresponding large contraction as it recedes from us. These are the conclusions that a single observer would make, and not those of two observers, as in usual explanation of the FitzGerald-Lorentz contraction. Weinstein came to same conclusions by plotting the exponent of the hyperbolic arc tangent rather than  the tangent because he did not go to the limit where $\dot{\rr}/c=\cos A$, which then defines the angle of parallelism. 

However, unlike the astronomical phenomenon of parallax, where the space constant $\kappa$ is so large and the parallax angle so small as to thwart all attempts to date at measuring a positive defect, the distortions predicted by \eqref{eq:approach} and \eqref{eq:recede} are actually more dramatic, precisely because of the finite speed of light. Since \[\pi-(\pi/2+A+B)<\pi/2-A=\phi,\] the defect is smaller than the complementary angle to $A$, known as the parallax angle in astronomy, $\phi$.  

The astronomical approximation, 
\[\tan\left(\pi/4-\Pi/2\right)=\tan\phi/2=\frac{1-\tan\Pi/2}{1+\tan\Pi/2}>r/2\kappa,\]
 which uses the smallness of $r/\kappa=-\ln\tan(\Pi/2)$ to give an upper bound of $2\tan\phi/2$, will not work here because $\tan(\Pi/2)$ can be quite small for values of $\dot{\rr}/c$ of the order unity that make the angle $A$ quite small. Moreover, since $A(=\pi/2-\phi)\le\Pi$, or $\phi>\pi/2-\Pi$, may place a larger lower bound on the parallax angle. Since $\phi$ is the upper bound of the defect, the latter may stand a greater chance of being measured. Although no astronomical lower bound for the parallelax angle has to date been found,  the non-Euclidean nature of light rays may be easier to access because the finite velocity of light is not a constraint on the hyperbolic measure of the velocity.

 \section{time invariant hyperbolic metric}

 To transform the exterior solution of the Schwarzschild metric  into the interior one \cite{Schwarz}, possessing  constant (negative) curvature, all that is necessary is to assume that the mass is not independent of the radius $\rr$. Then for objects in which the density $\varrho$ is essentially uniform, $m=\varrho \rr^3$, and introducing this into the Schwarzschild metric, renders it equivalent to the hyperbolic metric, \eqref{eq:lob}, with constant Gaussian curvature, $K=-4/\kappa^2$, where the absolute constant $\kappa=c/\sqrt{G\varrho}$, with $(G\varrho)^{-1/2}$ being the Newtonian free fall time. Gaussian curvature appears here as a relativistic effect, vanishing in the nonrelativistic limit as $c\rightarrow\infty$. Flatness cannot only be achieved in the limit of a vanishing density, but, also, in the case where relativistic effects are unimportant.
 
 The two cases of constant density and constant mass are distinguished by the different slopes of the curve of the velocity of rotation of galaxies as a function of their distance from the galatic center. For distances less than $\rr_c=2\times 10^{4}$ light years the curve rises with a constant positive slope. That means, if the centrifugal and gravitational forces just balance one another, the rotational velocity is proportional to the density, which remains essentially uniform. For distances greater than this amount, the curve slopes downward, where the rotational velocity is now proportional to the inverse square-root of the distance from the galatic center. This implies that the galatic mass is confined to a region whose volume has a radius of less than $\rr_c$, for once outside this volume it appears that the mass is independent of the radius, which is the Schwarzschild case. The crucial point is that although these two situations are physical different they both have the same zero curvature as predicted by the vanishing of the Ricci tensor.  Thus, it would appear, that  the Ricci tensor is not related to the presence of matter which casts doubt on the validity of the Einstein equations.  Moreover, it would also place in doubt the equivalence principle since centrifugal forces, corresponding to a hyperbolic space of constant negative curvature, are not equivalent to gravitational forces, that would be described by the Schwarzschild metric. 
 
 The transition from a system of constant mass to one of constant density is exemplified by the exterior and interior solutions to the Schwarzschild metric.  Schwarzschild sought his metric in the form:
 \begin{equation}
 ds^{2}=e^{\lambda}d\rr^{2}+\rr^{2}d\sigma^{2}-e^{\nu}c^{2}dt^{2}, \label{eq:ds}
 \end{equation}
 where $\lambda$ and $\nu$ are unknown functions of the radial coordinate $r$, and
 \[d\sigma^{2}=d\theta^{2}+\sin^{2}\theta d\phi^{2}.\]
  They are derived by Einstein's requirement that
 $R_{ik}$ is the contracted Ricci tensor vanishes. The Ricci tensor,
 \begin{equation}
 R_{ik}=\frac{\partial\Gamma^{l}_{il}}{\partial x^{k}}-\frac{\partial\Gamma^{l}_{ik}}{\partial x^l}+\Gamma_{il}^{r}\Gamma^{l}_{kr}-\Gamma_{ik}^{r}\Gamma_{rl}^{l}, \label{eq:Ricci}
 \end{equation}
 involves the Christoffel symbols,
 \begin{equation}
 \Gamma^{i}_{kl}=\half\frac{1}{g_{ii}}\left(\delta_{ik}\frac{\partial g_{ii}}{\partial x^{l}}+\delta_{il}\frac{\partial g_{ii}}{\partial x^{k}}-\delta_{kl}\frac{\partial g_{kk}}{\partial x^{i}}\right),\label{eq:Ch}
 \end{equation}
 and their derivatives
 It is clear from \eqref{eq:Ch} that only coordinate dependent elements of the metric tensor will influence the value of the Ricci tensor \eqref{eq:Ricci}, and Einstein's condition for the absence of mass, \eqref{eq:0}. Einstein built his \lq general\rq\ theory on the invariance of the space-time interval between two events, which can be considered as a natural generalization of the Minkowski four-dimensional invariance of the hyperbolic line element. First, this would be applicable to inertial systems, and second the criterion would not necessarily have any affect upon the Ricci tensor, since the squares of space or time intervals could enter with constant metric coefficients. 
 
 In particular, multiply the Schwarzschild line element \eqref{eq:ds} through by $e^{-\nu}$. Since the square of the time interval will not have any effect on the Ricci tensor, we are led to consider the spatial part
 \begin{equation}
d\Sigma^{2}=e^{\lambda-\nu}d\rr^{2}+e^{-\nu}\rr^{2}d\sigma^{2},\label{eq:Sigma}
 \end{equation}
 The components of the metric tensor are:
 \begin{equation}
 g_{11}=e^{\lambda-\nu}\;\;\;\;\;\;\;\;\;\;\;\; \mbox{and}\;\;\;\;\;\;\;\;\;\;\;\;\;\; g_{22}=\rr^{2}e^{-\nu}. \label{eq:g}
 \end{equation}
 The non-vanishing Christoffel symbols  are:
 \[
\Gamma_{11}^{1}=\half\left(\lambda^{\prime}-\nu^{\prime}\right), \Gamma_{12}^{2}=\frac{1}{\rr}-\half\nu^{\prime},   \Gamma_{22}^{1}=\left(\half\rr\nu^{\prime}-1\right)\rr e^{-\lambda}.\]
 These give the contracted parts of the Ricci tensor as:
 \begin{subequations}
 \begin{align}
R_{11}=\fourth\nu^{\prime}\lambda^{\prime}-\frac{1}{2\rr}\left(\nu^{\prime}+\lambda^{\prime}\right)-\half\nu^{\prime\prime}\nonumber\\
 R_{22}=\left\{\fourth \rr^{2}\nu^{\prime}\lambda^{\prime}-\half \rr\left(\lambda^{\prime}+\nu^{\prime}\right)-\half \rr^{2}\nu^{\prime\prime}\right\}e^{-\lambda}.\nonumber
 \end{align}
 \end{subequations}
 These terms do not vanish. What does vanish identically is the difference
  \begin{equation}
 R_{ii}-\half Rg_{ii}=0, \label{eq:0}
 \end{equation}
for $i=1,2$, where the scalar curvature is given by
\[R=g^{11} R_{11}+g^{22}R_{22}=\left\{\half\nu^{\prime}\lambda^{\prime}-\frac{1}{\rr}\left(\nu^{\prime}+\lambda^{\prime}\right)-\nu^{\prime\prime}\right\}.\]
Consequently, there is nothing invariant about any generalization of the Minkowskian line element, \eqref{eq:ds}, nor is there any validity to Einstein's criterion, $R_{ij}=0$, for the absence of mass. The expression \eqref{eq:0} vanishes independently of the absence, or presence, of mass, and independent of any explicit form for the parameters $\lambda$ and $\nu$.

The time-invariant metric form \eqref{eq:Sigma} becomes Lobachevskian,
\begin{equation}
d\Sigma^{2}=dr^{2}+\kappa^{2}\sinh^{2}(r/\kappa)d\sigma^{2} \label{eq:Lob}
\end{equation}
where $\rr$ and $r$ are related by \eqref{eq:tanh}. The coordinates $r$ and $\sigma$ (which in the plane $\theta=\pi/2$ reduces to $\phi$) are known as \lq semi-geodesic\rq\ coordinates. In the neighborhood of any point of the pseudosphere the metric form in semi-geodesic coordinates is of the form \eqref{eq:Lob}.

Taking the differential of \eqref{eq:tanh} gives the invariant hyperbolic line element, \eqref{eq:Lob}, as:
\begin{equation}
d\Sigma^{2}=\frac{d\rr^{2}}{(1-\rr^{2}/\kappa^{2})^{2}}+\frac{\rr^{2}d\sigma^{2}}{1-\rr^{2}/\kappa^{2}}. \label{eq:Lob-bis}
\end{equation}
 
 Moreover, the foregoing analysis shows that $g_{11} =1/g_{00}$, and this rules out Schwarzschild's inner solution for constants other than $A=0$ and $B=1$.

 The geometry of a surface $\rr=\rr_{1} =\mbox{constant}$ is the same as a sphere with Euclidean radius $\rr<\kappa$, which is also a hyperbolic sphere of radius
 \begin{equation}                                                                 
r_{1} = \kappa\int_{0}^{\rr_{1}/\kappa}\frac{dx}{1-x^{2}}=\kappa\tanh^{-1}(\rr_{1}/\kappa)
.\label{eq:H-length}
\end{equation}
As the Euclidean distance $\rr$ approaches the rim $\kappa$, the hyperbolic distance $r$ increases without limit.  Rulers  shrink as they approach the rim so that the radius $\rr_{1}$ actually has infinite hyperbolic length.
In contrast, Schwarzschild's line element \eqref{eq:ds}, where $e^{-\lambda}=1-\rr^{2}/\kappa^{2}$ for the inner solution, gives the radius as:
\begin{equation}
r_{1}= \int_{0}^{\rr_{1}/\kappa}\frac{dx}{1-x^{2}}=\kappa\sin^{-1}(\rr_{1}/\kappa).
\label{eq:H-length-bis}
\end{equation}
Now as $\rr_{1}\rightarrow\kappa$, $r_{1}\rightarrow\half\pi\kappa$, which requires a physical explanation.

The volume of the hyperbolic sphere is:
\begin{eqnarray}
V_{1}&=&\kappa^{3}\int_{0}^{2\pi}d\phi\int_{0}^{\pi}\sin^{2}\theta d\theta\int_{0}^{\rr_{1}/\kappa}
\frac{x^{2}dx}{(1-x^{2})^{2}}\nonumber\\
&=&2\pi\kappa^{3}\left[\frac{\rr_{1}/\kappa}{1-\rr_{1}^{2}/\kappa^{2}}-\tanh^{-1}(\rr_{1}/\kappa)\right]\nonumber\\
&=&2\pi\kappa^{3}\left\{\sinh\left(r_{1}/\kappa\right)\cosh\left(r_{1}/\kappa\right)-\frac{r_{1}}{\kappa}\right\}, \label{eq:V1}
\end{eqnarray}
which is infinite as $r_{1}$ increases beyond bound. This would be representative of a truly \lq open\rq\ universe. Rather, Schwarzschild's interior solution would give a volume
\begin{eqnarray}
V_{1}&=&\kappa^{3}\int_{0}^{2\pi}d\phi\int_{0}^{\pi}\sin^{2}\theta d\theta\int_{0}^{\rr_{1}/\kappa}\frac{x^{2}dx}{1-x^{2}}\nonumber\\
&=&2\pi\kappa^{3}\left[\sin^{-1}\left(\rr_{1}/\kappa\right)-\frac{\rr_{1}}{\kappa}\sqrt{1-\rr_{1}^{2}/\kappa^{2}}\right], \label{eq:V1-bis}
\end{eqnarray}
which has a maximum $\pi^{2}\kappa^{3}$ as $\rr_{1}\rightarrow\kappa$. A physical explanation of why a volume should be finite for a solution that supposedly describes an unbounded universe, is wanting.

\end{document}